\documentclass{aa}
\usepackage{txfonts}
\usepackage{color}
\usepackage{graphicx}
\usepackage{mathptmx}
\usepackage{longtable}
\usepackage{natbib}

\usepackage{t1enc}
\usepackage{times}
\usepackage[utf8]{inputenc}
\usepackage{color, colortbl}
\usepackage{graphicx}
\usepackage{footnote}
\usepackage[normalem]{ulem}
\usepackage{pifont}
\usepackage{lscape}
\usepackage{booktabs}
\usepackage{csquotes}
\definecolor{LightGray}{gray}{0.9}
\definecolor{LightGray1}{gray}{0.8}
\definecolor{LightGray2}{gray}{0.5}
\definecolor{pad}{rgb}{0.77,0.07,0.77}

\begin{document}

\title{The Catalogue of Cometary Orbits and their Dynamical Evolution}

\author{Małgorzata Królikowska\inst{\ref{inst1}} and Piotr A. Dybczyński\inst{\ref{inst2}}}

\institute{
Space Research Centre of Polish Academy of Sciences, Bartycka 18A, Warszawa, Poland, \email{mkr@cbk.waw.pl} \label{inst1}  \and Astronomical Observatory Institute, Faculty of Physics, A.Mickiewicz University, Słoneczna 36, Poznań, Poland, \email{dybol@amu.edu.pl}\label{inst2} 
}


\authorrunning{Kr\'olikowska \& Dybczyński}
\titlerunning{The CODE database.}

\offprints{M. Kr\'olikowska, \email{mkr@cbk.waw.pl}}

\date{Received xxxxx / Accepted yyyyyy}

\abstract{The new cometary catalogue containing data for almost 300 long-period comets that were discovered before 2018 is  announced (the CODE~catalogue). This is the first catalogue containing cometary orbits in five stages of their dynamical evolution, covering three successive passages through the perihelion, except the hyperbolic comets which are treated in a different manner.  For about 100 of these long-period comets, their non-gravitational orbits are given, and for a comparison also their orbits obtained while neglecting the existence of non-gravitational acceleration are included. For many of the presented comets different orbital solutions, based on the alternative force models or various subsets of positional data are additionally given. The preferred orbit is always clearly indicated for each comet.}

\keywords{catalogs/comets: general -- Oort Cloud}

\maketitle

\section{Introduction}\label{sec:intr}

The knowledge of orbits is crucial in many research areas related to the origin and the evolution of comets and their populations. In particular, the orbits of long-period comets (hereafter LPCs) are needed for studies of individual comet as well as the Oort Cloud dynamical evolution and their origin as a whole population; see for example \cite{rickman-2014} and \cite{Dones:2015} for reviews of the current unsolved problems. 

\begin{figure}
	\includegraphics[angle=270,width=8.8cm]{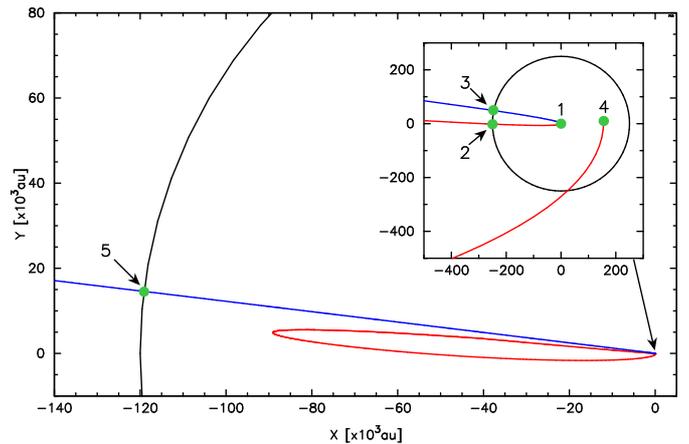} 
	\protect\caption{An example of a comet dynamical evolution described in CODE. This particular plot describes orbital changes of C/2002 T7 projected on its original orbit plane. Red line depicts the past motion of this comet while the blue one depicts its future evolution. Marked are five epochs (snapshots) when orbital elements are recorded: 1--osculating heliocentric orbit near the center of the observational interval (typically near the perihelion), 2--original barycentric orbit recorded in past at 250\,au from the Sun, 3--future barycentric orbit recorded in future at 250\,au from the Sun, 4--previous orbit, recorded at the previous perihelion, 5--next orbit, in this case recorded at the escape border at 120\,000\,au from the Sun, but for many other comets recorded in the next perihelion. } 
	\label{fig:orbits}
	
\end{figure}

We present the Catalogue of Cometary Orbits and their Dynamical Evolution (acronim: CODE catalogue). It is publicly available at the WebPage {\tt http://pad2.astro.amu.edu.pl/comets}. This is the first orbital catalogue showing orbital evolution of long-period comets during three consecutive perihelion passages:  previous -- observed -- future and covering their dynamical evolution over a period of typically 1--10\,million\,years. This is implemented by recording of five snapshots of osculating orbits: previous -- original -- observed -- future -- next; the positions of these snapshots along the trajectory are schematically shown on the example of orbit evolution of C/2012~T7 LINEAR presented in Fig.~\ref{fig:orbits}. 

At the moment, the CODE catalogue offers orbital solutions for almost a complete sample of comets discovered in the period 1885--2017  with an original semimajor axis greater than 10\,000\,au for a purely gravitational orbit. In this paper, we call these objects the Oort spike comets. Less than about twenty Oort spike comets with $q<3.1$\,au discovered in a period 2013--2017 are missing but will be added in the near future.

In May 2020, the CODE catalogue includes orbital solutions for 277~LPCs. In various dynamical aspects, most of these comets were described in a series of our previous papers; most recent are \citet[][KD2017]{kroli_dyb:2017} and \citet[][K2020]{kroli:2020};
some others (over 50~Oort spike comets discovered since 2012) will be analyzed in our next paper, currently in preparation.

In the next section, the procedure of the osculating orbit determination from the positional data is described, providing snapshot 1.  The method of obtaining snapshots~2 and 3, that is original and future orbits is described in Sect.~\ref{sect:ori_fut}.  The previous and next orbit calculations (snapshots~4 and 5) are sketched in Sect.~\ref{sec:previous_next}. Sect.~\ref{sec:CODE-general} is devoted to the description of the CODE database content and its Internet interface. An example of a multi-threaded approach to orbit determination in a several different variants is presented in Sect.~\ref{sec:C2002T7_LINEAR}, in the context of the comet C/2002~T7 with non-gravitational~effects difficult for modeling. Some statistics of the CODE~catalogue content is provided in Sect.~\ref{sec:statistics} and future plans are described in the last section.

\section{Osculating orbit determination; snapshot 1}\label{sec:osculating}

All osculating orbits of comets in the CODE Catalogue were determined from the positional data  using the same method of orbit determination and a model of the solar system as in our previous papers \citep[see for example][KD2010, K2014]{kroli-dyb:2010,krolikowska:2014}, including the procedure of data selection and weighting (Sect.~2 in KD2010). Only the aspects related to the non-gravitational (NG) acceleration are briefly discussed below. 

This homogeneity of methods and models used in the CODE catalogue is essential for statistical considerations of comet orbits, particularly in the investigations of a shape of the Oort spike maximum (that is the original $1/a$-distribution).

\subsection{Non-gravitational orbits }\label{sec:NG_models}

Over one third of LPCs presented in the CODE catalogue exhibit detectable NG~accelerations in their orbital motion. In the subsample of small perihelion distance comets ($q$<3.1\,au) NG~orbits are determined for over 60~per~cent of objects. For all these comets, we present a NG~orbit solution and additionally their purely gravitational orbit for the sake of comparison. 

To determine the NG-orbit we rely on the widely used formalism described by \citet{marsden-sek-ye:1973} where the three orbital components of the NG~acceleration acting on a comet nucleus are proportional to the $g(r)$-like function which is symmetric relative to a perihelion,

\begin{eqnarray}
F_{i}=A_{\rm i} \> g(r),& A_{\rm i}={\rm ~const~~for}\quad{\rm i}=1,2,3,\nonumber\\
 & \quad g(r)=\alpha(r/r_{0})^{m}[1+(r/r_{0})^{n}]^{k},\label{eq:g_r}
\end{eqnarray}

\noindent where $F_{1},\, F_{2},\, F_{3}$ represent radial, transverse, and normal components of the 
NG~acceleration, respectively, and the radial part is defined as positive outwards along the Sun-comet line. 

However, for some comets we applied different sets of coefficients $m,\, n,\, k$, the distance scale, $r_{0}$, and the normalization constant, $\alpha$ (fulfilling the condition: $g(1\,{\rm au})=1$). Generally, for comets with $q \leq 3.1$\,au  the standard values dedicated to a water sublimation were used, whereas for $q \geq 3.1$\,au we often took the values more adequate for a CO-sublimation (see Table~\ref{tab:gr-like_functions}). In a few cases, we managed to apply even a more dedicated approach, as in the case of comet C/2001~Q4 NEAT or C/2002~T7 LINEAR.  

Moreover, for some comets it was possible to determine also the fourth NG~parameter describing the shift of  a maximum of $g(r)$-function relative to the moment of perihelion passage \citep{yeomans-chodas:1989,sitarski:1994b}; see for example asymmetric NG~solutions for C/1956~R1~Arend-Roland, C/1959~Y1 Burnham, C/1990~K1~Levy, C/1996~E1~NEAT, C/1993~A1~Mueller, C/1998~P1~Williams, C/2002~T7 LINEAR, C/2007~W1~Boattini and others (see also K2020). However, not always such asymmetric solution can be considered as the preferred one; this is, for example, the case of C/2002~T7 (see Sect.~\ref{sec:C2002T7_LINEAR}).

It should be stressed that in all cases the NG~parameters  were obtained together with the osculating orbital elements in an iterative process of NG~orbit determination based on positional observations (more details in KD2010).

In the cases when the assumption of constant NG~parameters (see Eq.\ref{eq:g_r}) operating within the entire data arc is not adequate (due to an unexpected variable comet activity for any reason), it was worthwhile to use other dedicated approach, in particular, in the context of studying the past or future evolution of analyzed LPCs. Therefore in such a situation when NG~acceleration seems to be significantly different before and after perihelion passage, we also present separate orbit solutions basing on the observations taken before the perihelion passage and separately after the perihelion passage, or give a solution based on data taken only at large heliocentric distances;  see for example different orbital solutions for: C/1993~A1 Mueller, C/2001~Q4 NEAT, C/2002~T7 LINEAR, C/2003~K4 LINEAR, C/2007~W1 Boattini, C/2008~A1 McNaught, and others.  

In the sample of LPCs with small perihelion distances, especially for those with $q\le 1$\,au, a sudden outbursts and/or split event were often observed \citep{sekanina:2019}. Then, we derive the orbit (gravitational or NG~orbit) on the basis of data restricted to period before the detected sudden activity in a given comet, see for example C/1999~S4 LINEAR, C/2010~X1 Elenin, C/2002~O4 H{\"o}nig, C/2002~O7~LINEAR, C/2012~S1 ISON, and others.

As a result of different possible force models and alternative  approaches to available positional data, a set of various orbital solutions is presented for many comets in the CODE Catalogue, with an indication of the preferred orbit; for more details see K2020. The case of C/2002~T7 is the representative example of various solutions presented in the CODE catalogue (see Sect.~\ref{sec:C2002T7_LINEAR}). In total, at the moment of this writing, we present 508~different orbital solutions for 277~individual LPCs.

\begin{center}
	\begin{table}
		\caption{\label{tab:gr-like_functions}Parameters generally used in Eq.~\ref{eq:g_r} in CODE Catalogue}
		
		\setlength{\tabcolsep}{5.0pt} 
		\begin{tabular}{lllll}
			\hline 
			\\
			\multicolumn{5}{c}{standard g(r)-function (water sublimation) }\\
			\\
			$\alpha$    & $r_0$ & $m$      & $n$   & $k$           \\
			0.1113      & 2.808 & $-$2.15  & 5.093 & $-$4.6142     \\
			\\
			\multicolumn{5}{c}{g(r)-like function (CO sublimation)} \\
			\\
			$\alpha$    & $r_0$ & $m$      & $n$   & $k$           \\
			0.01003     & 10.0  & $-$2.0   & 3.0   & $-$2.6        \\
			\hline
		\end{tabular}
	\end{table}
\end{center}

\section{Original and future cometary orbits; snapshots 2 \& 3}\label{sect:ori_fut}
 
For the past and future dynamical evolution of a given comet a swarm of 5001\, virtual comets (hereafter VCs), including the nominal orbit were constructed according to the Monte Carlo method proposed by \citet{sitarski:1998}. This  approach allowed us to determine the uncertainties of all orbital elements at any epoch covered by a numerical integration to the previous perihelion passage as well as to the subsequent perihelion passage.
 
Next, the dynamical evolution of each swarm of VCs  was numerically followed backwards and forwards in time until each VC reached 250\,au from the Sun, that is, a distance where planetary perturbations are already negligible. At this stage we also switch from a heliocentric to a barycentric reference frame. These swarms of orbits are called original (snapshot~2) and future (snapshot~3) and are available for download as described in Sect.~\ref{sec:CODE-browse}.

C/2002~T7 LINEAR is an interesting example of various approaches to determining its orbit \citep[][K2020, and Sect.~\ref{sec:C2002T7_LINEAR}]{kroli-dyb:2012}.  The trajectory of this comet is shown in Fig.~\ref{fig:orbits} and the position of the observed perihelion is indicated in an inset by a green point with a label '1', while positions  corresponding to the original and future snapshots are given by green points with labels '2' and '3', respectively. Due to a small perihelion distance of 0.615\,au the perihelion point merges with the position of the Sun in this plot.

\section{Previous and next cometary orbits; snapshots 4 \& 5}\label{sec:previous_next}

Original and future swarms of LPCs' orbits were next followed numerically taking into account Galactic and stellar perturbations. We use models and methods described in detail in \citet{dyb-berski:2015}. In short: we numerically integrate each virtual comet with the Sun and several hundreds of nearby stars under the overall Galactic potential.

Our list of potential stellar perturbers includes at present 643 stars or stellar systems that can appear closer than 4\,pc from the Sun during the studied time intervals (typically 10\,Myr to the past or to the future). The detailed description of this list and of the publicly available database of stellar perturbers\footnote{\tt  https://pad2.astro.amu.edu.pl/stars/} can be found in \citet{wys-dyb-poli:2020}.

To obtain snapshots 4 and 5, we generally stop our calculation at the previous (when numerically integrating an orbit backward in time) or next perihelion (when a comet orbit is integrated forward in time). However, for hyperbolic or extremely elongated elliptic orbits in a swarm of VCs, we apply the so-called {\it escape limit} of 120\,000~au; see snapshot~5 in Fig.~\ref{fig:orbits}. The final orbits obtained from these calculations, we call previous (snapshot~4) and next (snapshot~5). In the catalogue, we call a comet (or more precisely each individual VC) as returning [R] if it goes (backward or forward in time) no further than 120\,000~au from the Sun.  All other VC are generally named escaping [E]; however, in this group we separately count a number of hyperbolic [H] orbits. 

In many cases a variety of the dynamical behavior among VCs within a particular swarm of orbits enforced the use of individual rules of stopping the numerical integration. We decided to treat each swarm of orbits as uniformly as possible, thus, we distinguish between the following three cases.

\begin{enumerate}
	\item When all VCs of a particular comet were returning then all of them were stopped in two ways: 
	
	\begin{itemize}
		\item at the previous/next perihelion (a basic variant),
		\item simultaneously, at the moment when the nominal VC reached previous/next perihelion (a synchronous variant).
	\end{itemize}

\item When all VCs were escaping then the calculation was stopped synchronously, at the moment when the fastest VC cross the escape limit (usually being very close to a heliocentric distance of 120\,000~au).

\item When a swarm of VCs consists of both returning and escaping VCs we also decided to stopped them in two ways: 

\begin{itemize}
	\item the returning part was stopped at the previous/next perihelion and the rest (escaping ones) when the fastest escaping VC crossed the escape limit (mixed variant).
	\item all VCs (both returning and escaping) were stopped at the moment when the fastest VC crossed the escape limit (synchronous variant).
\end{itemize}
\end{enumerate}

The synchronous variant is used to examine more homogeneously constructed orbital element distributions, that is  when all VCs were stopped exactly at the same epoch.
	
In the CODE catalogue, we describe parameters of the previous and/or next orbit of a comet using statistics of basic or synchronous variant depending on an individual swarm structure. It means, that for previous and next orbits, we present:

\begin{itemize}
	\item a number of returning [R], escaping [E], and hyperbolic [H] (among the escaping) VCs in a swarm. We mark in which of these three parts of VCs, the nominal orbit can be found,
	\item a reciprocal of the semimajor axis,
	\item an aphelion distance,
	\item a time interval to the previous/next perihelion. 
\end{itemize}

The last three parameters are presented in two variants. In the case, when their distribution in the swarm is close to the Gaussian, we present the mean value and  its standard deviation. If the distribution of a particular parameter is far from being  a normal one, we present three deciles: the first (10 per cent of distribution), the fifth (median) and the ninth (90 per cent). We additionally present a statistics of the previous/next perihelion distance, showing a percentage of its values smaller than 10~au, being between 10 and 20\,au, and greater than 20\,au. For previous orbits, we interpret these ranges of values as defining whether the particular comet is dynamically old, have uncertain status, or is dynamically new, respectively (more details in KD2017).

How all the above rules works are shown on an example of C/2002~T7 in Fig.~\ref{fig:orbits}. According to the preferred solution, this comet passed its previous perihelion at a distance of about 150\,au, see green point labeled by '4'. However, this Oort spike comet escapes from the planetary zone on a hyperbolic orbit. Therefore, the snapshot '5' was taken at the moment when this comet reached the 120\,000\,au from the Sun in our dynamical calculations of its future motion (now this comet is less than 40\,au from the Sun).

\section{CODE database usage}\label{sec:CODE-general}

There are two main views of the CODE database interface, called 'Browse' and 'Search'. The first one allows the user to browse across all orbital solutions available in the database, including osculating, original, future, previous, and next orbits described in the preceding sections.

The 'Search' screen allows for searching among all the data narrowing almost all parameter intervals. Below, we describe in short both these functionalities.

An anonymous user have an access to a slightly narrowed subset of data, for example objects currently under an additional study might be partially embargoed. Logged users can access some additional objects/orbits. Logging credentials can be obtained by an e-mail from the authors on the circumstantiated request, for example in case of a research cooperation on particular objects.

\subsection{Browsing the CODE database}\label{sec:CODE-browse}

'Browse' is a default view of the CODE database, available immediately after opening its Internet main page in a browser. It shows all comets with their designations, names, observational material characteristics and osculating orbit elements (presented in lowered precision). In this view, we present only one, preferred orbit for each comet. The user can narrow the visible list of comets by using a filtering box (given at the upper left corner). Any string entered in this box will be automatically matched against a comet designation and/or its name. Moreover, clicking on any column headings will sort all the table according to the pointed parameter.

Clicking on the comet designation presented in the leftmost column allows to enter a 'comet presentation screen'. In its upper part, a comet description is available and a clickable list of orbital solutions is shown (the preferred one is presented by default and named in green).

For each orbital solution five cards are available, describing in a more detailed manner and with the greater precision five orbital snapshots: osculating, original, future, previous, and next, presented from the left to the right in a chronological order. In most cases an additional graphics are also presented.

As it was mentioned in Sect.~\ref{sect:ori_fut}, we constructed a swarm of 5001 'virtual comets', including the nominal osculating orbit for each of orbital solutions in the database. By propagating in time all these VCs to the previous or next perihelion, we can estimate the uncertainties for all presented orbital parameters. For those who intend to perform similar investigations, but for example on much longer time scales, we offer downloading a text file consisting of all  5001~VCs at the stage of original or future orbits; see the appropriate cards with the 'original' or 'future' orbit. 

\subsection{Searching the CODE database}\label{sec:CODE-search}

After entering the 'Search' view, the user is offered with a possibility of choosing a subset of orbital solutions available in CODE by narrowing intervals of almost all parameters. By default, the 'Search' screen starts the search among osculating orbits, but it can be switched to the other four types of snapshots. By checking an appropriate box (given on the right), one can limit the search to the preferred orbits only.

Orbital elements are selected by defining an interval of interest and a simple check of correctness of the entered data is performed dynamically. A special treatment is performed for four elements: a perihelion and aphelion distance, an eccentricity, and a semimajor axis. Since these elements are not independent, the user might enter the preferred interval of any two of them and the rest is automatically blocked. When selecting an interval of the perihelion passage epoch one can use different limitations for: a year, a year+month or a year+month+day. 
In the lower part of the screen a selection basing on the observational data characteristics or an orbital solution type can be done.

Searching among 'previous' and 'next' orbits is performed with a little different form, adapted to the information presented for these types of orbits.

After clicking the 'Submit' button, the user obtains a list of matching orbital solutions in a form identical with the 'Browse' screen (which furthermore can be filtered by a comet designation or name). There is only one important difference: the search results can be downloaded as a simple text file, with all the data presented in a full precision together with their uncertainties. 

\begin{figure}
	\centering
	\includegraphics[width=0.38\textwidth, angle=270]{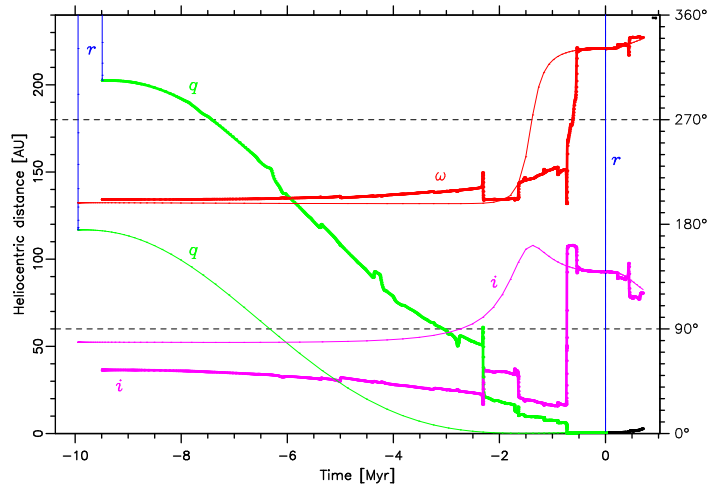}
	\caption{Dynamical evolution of the nominal orbit of C/2002~T7 LINEAR ('d6' solution) under the simultaneous Galactic and stellar perturbations. The left vertical axis describes a heliocentric distance plot (thin blue lines) and a perihelion distance (green lines, that are changed into black if $e\geq1.0$). The right vertical axis describes the angular elements (with respect to the Galactic disk plane): an inclination (fuchsia line) and an argument of perihelion (red line). Thin lines  describe the dynamics when all stellar perturbations are omitted.}
	\label{fig:2002t7 nominal_evolution}
\end{figure}

\section{C/2002~T7 -- the example case description}\label{sec:C2002T7_LINEAR}

In the CODE catalogue we divided all comets into four categories on the basis of the exposure of NG~effects in positional data fitting. To which category we have classified a given comet can be found in a card with a purely gravitational osculating orbit.  Details of this approach are described in Sect.~5 of K2020.

C/2002~T7 belongs to the group of comets with NG~effects strongly manifested in positional data fitting.
At this moment less than 20~LPCs are included to this group in the CODE catalogue (see also KD2020).

Additionally, this comet was discovered at a large heliocentric distance and were followed for a long time. 
Thus in this case, a satisfactory fit to the data for the NG~model of motion (based on $g(r)$-formula adequate for water ice sublimation) can be obtained without great loss of the orbital accuracy when positional data around perihelion are neglected or when we try to derive the orbit from data taken before perihelion for past dynamical evolution (or data taken only after perihelion for future dynamical studies). 

For this comet even an asymmetric NG~model do not fit the whole data set satisfactorily, because some trends in observed minus calculated (O$-$C) residuals in right ascension and/or declination are visible (see the database). The better data fitting is obtained when positional data around perihelion are neglected and such a NG~model based on DIST  type of data (solution 'd6', see the database or K2020) was chosen as the preferred one when we would like to have a general orbital solution for this comet. 

In Fig.~\ref{fig:2002t7 nominal_evolution} we present the past and future evolution of the nominal orbit of C/2007~T7 (the preferred solution d6) using the dynamical model adopted for the CODE~catalogue. The thick lines describe the past (to the previous perihelion) and future (to the escape limit crossing) orbit evolution under the simultaneous Galactic and stellar perturbations. The list of the potential stellar perturbers collected for the CODE catalogue calculations is described in \cite{wys-dyb-poli:2020}. For the sake of comparison an evolution with all stellar perturbations omitted is also shown (thin lines). 

It is easy to note that the previous perihelion is well outside the planetary zone, so we classified this comet as the dynamically new one.

However, there are also other satisfactory solutions based on pre-perihelion data. Fortunately, all NG~models representing different approaches to data give a very similar original orbit for this comet with $1/a_{\rm ori}$ in the range from 17.75 in units of $10^{-6}$\,au$^{-1}$ (NG~solution: p6, based on pre-perihelion data) to 28.93 in the same units ( asymmetric NG~solution: 'n6', entire data set). For all of them, we conclude that this comet is a dynamically new one by obtaining its previous perihelion distance (snaphot~4). The preferred model (d6) with $21.65\pm 0.86$ is in the middle of this range.


In the~CODE catalogue we also include one solution based on $g(r)$-like function describing CO-sublimation. Such a solution (named 'pc' and based on pre-perihelion data only) gives a significantly different value of  $1/a_{\rm ori}=53.21\pm3.63$ in units of $10^{-6}$\,au$^{-1}$. This solution leads to the result that C/2002 T7 was at a distance of about 5-7\,au from the Sun at the previous perihelion passage. In this case, we should classify this object as a dynamically old comet. However, the situation that a comet being dynamically old is still sublimating mainly CO is rather unlikely.

\section{Some statistics}\label{sec:statistics}

\begin{figure}
	\centering
	\includegraphics[width=1.0\columnwidth]{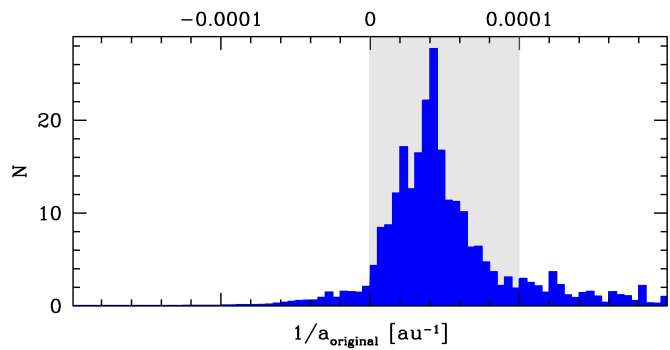}
	\caption{Distribution of $1/a_{\rm ori}$ based on the preferred orbits for the whole sample of LPCs available in CODE~catalogue, where about one third has NG~solutions. The light gray vertical band indicates the region occupied by Oort spike comets.}
	\label{fig:aori_distribution}
\end{figure}

At the moment, the CODE catalogue contains data for 277~Oort spike comets. Since for some of them different orbital solutions are presented in the database, we show a total of 508~different orbits; however, the preferred orbit is always indicated for each comet. In the case where it was possible to obtain NG~orbit, two types of orbits are always given for such comet: NG and purely gravitational. This approach allows to independently examine this sample of comets from various points of view; for example, to study the impact of NG~effects on the change of the osculating orbit, etc. Many similar statistical analyses were previously done by us for smaller samples of Oort spike comets. Here, we present one result of such analysis that can be performed based on the original orbits for solutions indicated as preferred in this database. 

Fig.\ref{fig:aori_distribution} displays the $1/a_{\rm ori}$ distribution of the whole sample of all comets given at this moment in the CODE catalogue. This histogram was constructed using the full swarms of preferred orbits, see \citet{dyb-kroli:2016} for a detailed description of this technique. Thus, this distribution is a sum of individual Gaussian distributions of $1/a_{\rm ori}$, where the dispersion of each Gaussian reflects the individual uncertainties of $1/a_{\rm ori}$ of each comet. All of these individual distributions are also available in the database (see Sect.\ref{sec:CODE-browse}).

\section{Future plans}\label{sec:conclusion}

In the near future, we plan to investigate Oort spike comets with $q<3.1$\,au discovered in the period 2013--2017 (about 20 objects). When supplemented with these objects, the CODE catalogue will contain a complete sample of Oort spike comets\footnote{defined by its pure gravitational solutions} from the period 1885-2017, with an exception of a few comets discovered long ago, requiring a special and time consuming approach, including completing positional data from highly dispersed sources, see the case of C/1890 F1 \citep{kroli-dyb:2016}.

When the observations are finished and the data set is fixed then the  osculating orbit(s) of a particular comet can be treated as 'definitive', except the situations when we deal with the original old positional data based on old stellar catalogues. An example of such a recent correction to the definitive comet orbit obtained long ago one can found in \cite{kroli-dyb:2016}.

Slightly different situation is in a case of original and future orbits. Once the existence of the hypothetical planet nine is unambiguously documented, it cannot be ruled out that some of original and future orbits may experience some, probably minor/slight, modifications.

However, the previous and next orbit calculations strongly depend on our detailed knowledge on the solar system Galactic neighbourhood. Our current list of stellar perturbers \citep{wys-dyb-poli:2020}, based on the {\it Gaia} DR2 catalogue \citep{Gaia-DR2:2018}, still contains several problematic objects which have to be examined further when new data will be available. This uncertain perturbers can affect both past and future cometary motion, so previous and next orbits of many LPCs included in the CODE catalogue might be changed in the future releases. 

Therefore, this catalogue provides orbital data based on the current state of our knowledge about the Solar System and its surroundings.

We also plan to enrich the database interface with more graphical capabilities, including an application that will allow to obtain a picture of the dynamical evolution of orbits under Galactic and stellar perturbations over a time scale of about $\pm 10$\,million\,yr.

\begin{acknowledgements}
This research was partially supported by the project 2015/17/B/ST9/01790 founded by the National Science Centre in Poland.
\end{acknowledgements}

\bibliographystyle{aa}
\bibliography{PAD30.bib}

\label{lastpage}

\end{document}